\documentclass[a4paper,11pt]{karticle}
\usepackage{graphicx}

\renewcommand{\thesection}{\S\arabic{section}}

\title
{
Simulation of a Dripping Faucet
}

\author
{ 
Nobuko {\sc Fuchikami}\footnote{E-mail: fuchi@phys.metro-u.ac.jp} , Shunya {\sc Ishioka}$^{1}$ and Ken {\sc Kiyono}
}

\date
{
(Received October 15, 1998)
}

\begin{document}
\sloppy
\maketitle
\begin{center}
\textit{
Department of Physics, Tokyo Metropolitan University, Tokyo 192-0397\\
$^1$Department of Information Science, Kanagawa University, Kanagawa 259-1293
}
\end{center}

\begin{abstract}
A dripping faucet system is simulated. 
We numerically show that a hanging drop generally has many equilibrium 
shapes but only one among them is stable. 
By taking a stable equilibrium shape as an initial state, 
a simulation of dynamics is done, 
for which we present a new algorithm based on Lagrangian description.
The shape of a drop 
falling from a faucet obtained by the present algorithm agrees
quite well with experimental observations. 
Long-term behavior  
of the simulation can reproduce period-one, period-two, intermittent 
and chaotic oscillations widely observed in experiments. 
Possible routes to chaos are discussed.  
\end{abstract}

{\footnotesize
\begin{center}
\end{center}
}

\baselineskip 5.5mm

%%%%%%%%%%%%%%%%%%%%%%%%%%%%%%%%%%%%%%%%%%%%%%%%%%%%%%%%%%%
%%%  1.Introduction                                     %%%
%%%%%%%%%%%%%%%%%%%%%%%%%%%%%%%%%%%%%%%%%%%%%%%%%%%%%%%%%%%
\section{Introduction}
\newcommand{\lsim}{\begin{minipage}{12pt}
\vspace{0pt}$
\,\stackrel{\textstyle <}{\sim}$\end{minipage}}
\newcommand{\gsim}{\begin{minipage}{12pt}
\vspace{0pt}$
\,\stackrel{\textstyle >}{\sim}$\end{minipage}}

The formation of drops is an intriguing phenomenon
widely observed in everyday life.
Although scientific researches on this subject date back to the seventeenth
century\cite{eggers97}, great progress has been achieved only recently, mainly
in detailed studies on the behavior of drops near the breakup point. 
Breakup of a drop is a critical phenomenon 
corresponding to a singularity of a nonlinear partial differential equation 
obeyed by the fluid with free surface.
Refined experiments to observe drops falling from nozzles 
for various viscosities 
provide much knowledge on this critical 
phenomenon\cite{pss90,sbn94}. 
Eggers proposed a scaling theory which is universally
applicable for axisymmetric drops with finite viscosity\cite{eggers93}. 
Numerical solutions of Navier-Stokes equations have been obtained
with high precision for viscid\cite{eggers97,ed94,eggers97a} 
and inviscid\cite{schulkes94}
fluids, and well reproduce the observed shape of drops. 

Another interesting aspect of the drop formation is 
the long-term behavior of a dripping faucet as a chaotic dynamical system.
Since Shaw's pioneering work\cite{shaw84}
it has been confirmed in many
experiments that dripping time intervals exhibit 
various types of periodic and chaotic oscillations including 
intermittency and hysteresis\cite{mpss85,ws89,dh91,sgp94}. 
Drastic change of an attractor is induced  
%Attractor changes drastically 
by a small variation of 
the flow rate, which is a main control parameter of the system.
However, theoretical progress in this direction is not yet satisfactory.
A simple ``mass-on-a-spring'' model has been proposed and 
an analog simulation of this model was reported to reproduce, 
in a qualitative sense, some experimental return maps\cite{shaw84,mpss85}.
A numerical simulation based on a stochastic model was presented recently,
in which an Ising-like system was used\cite{op93}.
Since the latter model ignores kinetic terms, 
it might be applicable only for very 
small diameter of faucet($\lsim$ 0.5mm). 
In contrast, the mass-spring model
corresponds to relatively large faucets, where
the oscillation of the center of mass governs the basic dynamics. 
However, to explain the complex behavior observed in experiments 
systematically, 
information on the model parameters (for example, the spring constant) 
is essential, which is, unfortunately, not easily obtained.
Both theoretical studies do not aim at investigating the realistic shape 
of drops, focusing on the long-term behavior of the dripping time intervals.
%We consider that to reproduce the shape reasonably is also important to know how to make a simplified model. 

In our views, the reproduction of the shape of a drop is crucial 
in general understanding of the physics of the phenomenon.
Indeed, our detailed simulation turns out to be successful in this respect. 
Even when one attempts to analyze the phenomenon using a simplified model, 
the knowledge of the drop shape is necessary for the correct choice of 
parameters.  
Our first goal is a long-term simulation using an algorithm which can
also simulate the shape of drops reliably. After that, 
on the basis of detailed analysis of numerical data thus obtained, we are 
going to construct 
an improved mass-spring model which we expect reveals essential
features of this complex system.

In this paper, we present a new algorithm for simulating a dripping 
faucet based on 
Lagrangian description instead of Eulerian one like Navier-Stokes
equations. We decompose a drop into many 
parts (at most 300 $\sim$ 400 disks or less) and 
describe the dynamics in terms of time evolution 
equations obeyed by each part under the influence of gravity, 
surface tension and viscosity.   
After we observe that our algorithm can well reproduce 
shape of a drop at various stages of evolution, 
we proceed to a long-term simulation 
in which the growth and fission of a drop take place many times
under a constant increase of the mass of the fluid.
It will be shown that
time series of the dripping time intervals reproduce  various types of 
motions including period-one, period-two, 
intermittent oscillations etc. besides chaotic ones,  
as observed in many experiments.
A bifurcation diagram will be presented which agrees well with a recent
experiment. 

In \S 2, we derive static shapes which are used as an initial condition
for solving dynamical equations. 
In \S 3, equations of motion in Lagrangian description
is presented. Computational results and experimental data are
compared in \S 4. In Appendix A, we describe a variational algorithm to
examine stability of static solutions. 
Our algorithm of simulation of dynamics is presented in Appendix B.

%%%%%%%%%%%%%%%%%%%%%%%%%%%%%%%%%%%%%%%%%%%%%%%%%%%%%%%%%
%%%  2. Equilibrium states                            %%%
%%%%%%%%%%%%%%%%%%%%%%%%%%%%%%%%%%%%%%%%%%%%%%%%%%%%%%%%%
\section{Equilibrium States}
\label{sec:heikou}

We first derive static equilibrium shapes of a pendant drop where 
the gravitational and the surface tension forces balance each other.
There is a maximum (i.e., critical) volume $V_{\rm c}$ for
the stable state when the radius $a$ of the faucet is fixed.
We will see below that 
there exist in general several equilibrium shapes when a volume
smaller than $V_{\rm c}$ is given. 
As suggested by Padday and Pitt\cite{pp73}, only one among them  
is stable and realized, which we will show
numerically.

The force balance equation of the drop is
%\begin{eqnarray}
\begin{equation}
P = \rho g z,
\label{eq:heikou} 
\end{equation}
where $z$ is the vertical coordinate (the positive
direction is defined to be downward), $P$ the pressure. 
The density $\rho$ is assumed to be
constant throughout the fluid.
The pressure (difference between the inside and the outside of the
interface) is expressed in terms of the principal radii of 
curvature of the drop surface, $R_1$, $R_2$ as
\begin{equation}  
P = \Gamma \left(\displaystyle\frac{1}{R_1} + 
\frac{1}{R_2} \right)\,, 
\label{eq:curvature}
\end{equation}
where $\Gamma$ is the surface tension.
Choosing the length, mass, and pressure units as
\begin{equation}
l_0 \equiv \sqrt{\Gamma / \rho g}\,, \quad
m_0 \equiv \rho l_0^3\,, \quad
P_0 \equiv \sqrt{\rho g \Gamma}\,, 
\label{eq:unit} 
\end{equation}
%$$
%l_0 \equiv \sqrt{\Gamma / \rho g}\,, \quad
%m_0 \equiv \rho l_0^3\,, \quad
%P_0 \equiv \sqrt{\rho g \Gamma}\,, 
%$$
( $l_0 =0.27$cm, $m_0 = 0.020$g and $P_0 = $270dyn/cm$^2$
for water at 20 $^{\circ}$C,)
we can set $g= \rho = \Gamma =1$.
 
\begin{figure}[p]
%\figureheight{1cm}
	\begin{center}
	\includegraphics[width=.4\linewidth]{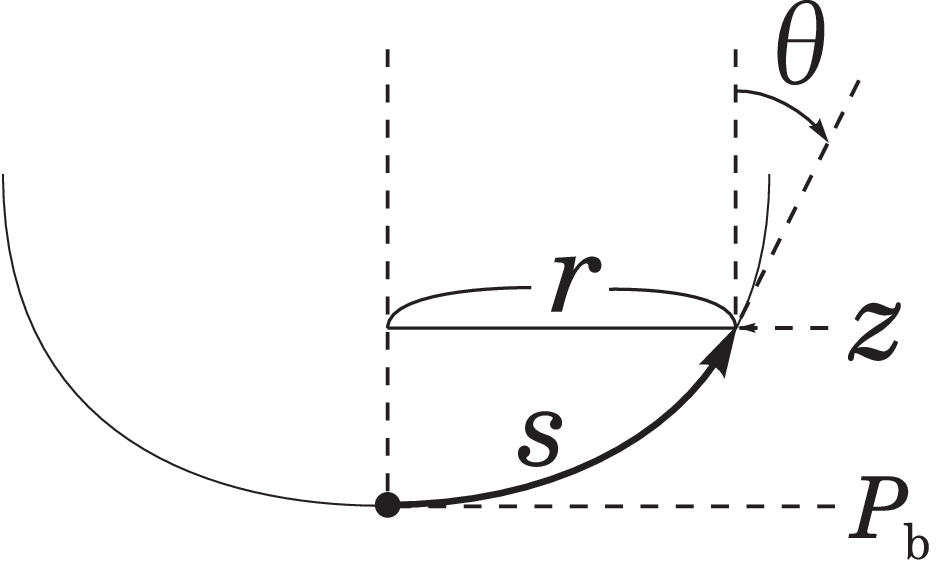}
	\end{center}
\caption{Definition of variables.}
\label{fig:var_def}
\vspace{1cm}
\begin{center}
\end{center}
\end{figure}

For an axisymmetric drop, the radii of curvatures are given by
\begin{eqnarray}
\displaystyle\frac{1}{R_1} &=& -\frac{{\rm d}\theta}{{\rm d}s}\,, \\
\frac{1}{R_2} &=& \frac{\cos\theta}{r}\,, 
\end{eqnarray}
where the variables $s$, $r$ and $\theta$ are defined in Fig.~\ref{fig:var_def}.
The shape of the drop is thus obtained by integrating a set
of ODE's:
\begin{equation}
\left.
\begin{array}{lll}
\displaystyle\frac{{\rm d}r}{{\rm d}s} &=& \sin\theta, \\
&&\\
\displaystyle\frac{{\rm d}z}{{\rm d}s} &=& -\cos\theta, \\
&&\\
\displaystyle\frac{{\rm d}\theta}{{\rm d}s} &=& \displaystyle\frac{\cos\theta}{r} -z,
\end{array}
\right\}
\label{eq:heikou1}
\end{equation}
with an  `initial' condition: $r(0)=0$, $z(0)=P_{\rm b}$, $\theta(0)=\pi /2$ 
%with an 'initial' condition: $r(0)=0$, $z(0)=P_{\rm b}$, $\theta(0)=\pi /2$ 
at $s=0$, where
$P_{\rm b}$ is the pressure at the bottom of the drop.

For a given value of $P_{\rm b}$, the solution of eq.~(\ref{eq:heikou1}) 
is unique. 
However, we need a boundary condition at the top of the drop to
determine the upper limit of integration.
%05jan99
Here it should be noted that eqs.~(\ref{eq:heikou}) 
and (\ref{eq:curvature}) describe only the force balance, namely,  
they correspond to an equilibrium condition but not to any stability condition.
We therefore have to select stable shapes out of the solutions of 
eq.~(\ref{eq:heikou1}).
It is generally a difficult mathematical problem to examine whether an 
equilibrium state is stable or not when the system energy has not a lower 
limit, just like the present hanging drop. 
To the best of our knowledge, the existence of a stable shape and its 
uniqueness have not been proved.
Padday and Pitt investigated stable shapes of drops under various conditions, 
assuming implicitly the uniqueness of the stable solution. 
In this paper, we numerically confirm that when there is one or more 
solutions of eq.~(\ref{eq:heikou1}), 
all except one are unstable. 
The rest is stable under any infinitesimal deformation, 
which we cannot prove but we conjecture.

%%% 2.1 %%%%%%%%%%%%%%%%%%%%%%%%%%%%%%%%%%%%%%%%%%%%%%
\subsection{Drop hanging from a faucet}

If we assume that the inner surface of the faucet is fully wetted
and the outer surface is not wetted at all, the boundary condition is 
volume-radius limited\cite{pp73}. That is, 
the shape of the drop is obtained
by integrating the above equations up to $r=a$, where $a$ is 
the inner diameter of the faucet. Solution $r$ of eq.~(\ref{eq:heikou1})
oscillates as $s$ (and hence $z$) varies, which leads to, in general,  
many equilibrium shapes for fixed boundary condition.
We can show, however, that at most one among them (the shortest shape)
is stable.

\begin{figure}[p]
%\figureheight{1cm}
	\begin{center}
	\includegraphics[width=1\linewidth]{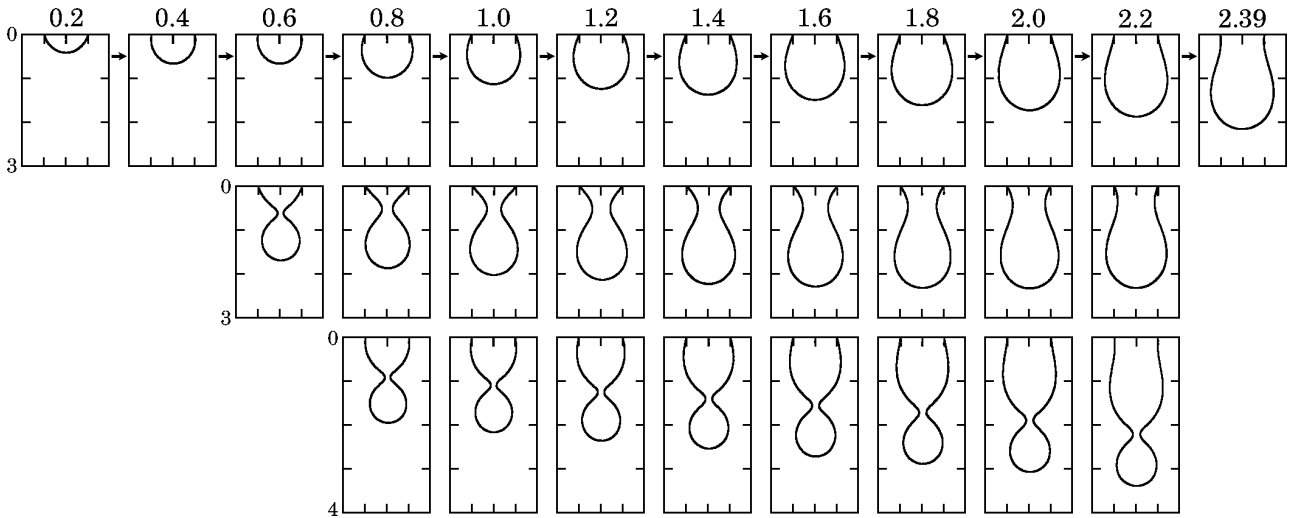}
	\end{center}
\caption{Drop hanging from the faucet. $a=0.5$.
Various equilibrium shapes for a fixed volume of the drop. 
Among them, only one (the top of each column)
is stable. As the volume is increased, the shape changes as 
indicated by arrows. The maximum volume is 
%%%%%$V=2.386$.}
$V=2.39$.}
\label{fig:shape_v05}
\vspace{1cm}
\begin{center}
\end{center}
\end{figure}

\begin{figure}[p]
%\figureheight{1cm}
	\begin{center}
	\includegraphics[width=.5\linewidth]{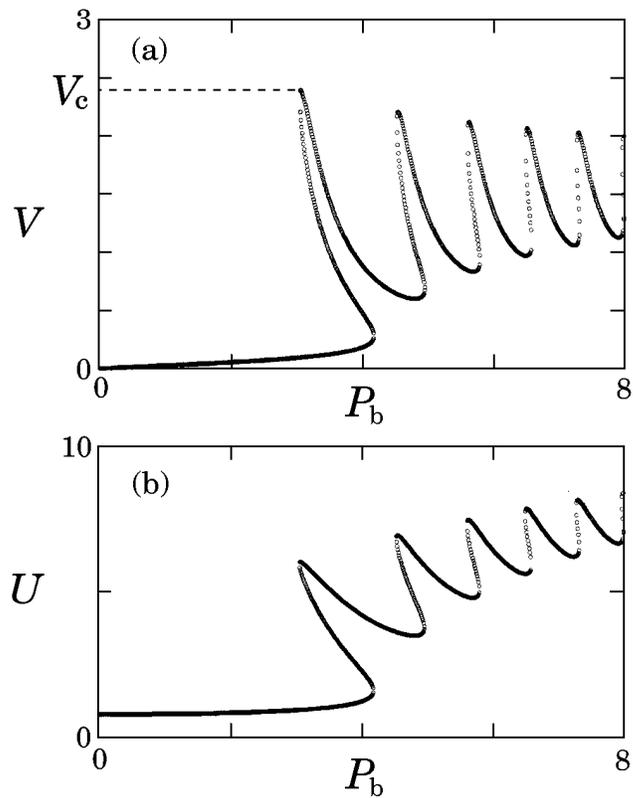}
	\end{center}
\caption{Plot of the volume $V$ and the energy $U$ against the pressure at 
the bottom, $P_{\rm b}$ for a drop hanging from the faucet. 
%%%%%$V_{\rm c}=2.386$ 
$V_{\rm c}\approx 2.39$ 
is the critical volume.
The radius of the faucet $a=0.5$.}
\label{fig:vu_pb05}
\vspace{1cm}
\begin{center}
\end{center}
\end{figure}

In the following discussion, we fix at $a=0.5$.
Figure \ref{fig:shape_v05} shows various profiles of drop obtained from 
eq.~(\ref{eq:heikou1}).
The volume of drop, $V$ and its potential energy, $U$
are plotted against the bottom pressure, $P_{\rm b}$ in 
Figs.~\ref{fig:vu_pb05}(a) and \ref{fig:vu_pb05}(b), respectively.
The energy $U$ is the sum of the gravitational energy $U_g$
and the energy of surface tension $U_{\Gamma}$: $U=U_{g}+ U_{\Gamma}$.
These quantities are calculated from solutions of eq.~(\ref{eq:heikou1}).
Figure \ref{fig:vu_pb05}(a) indicates that, as a function of $V$,  
$P_{\rm b}$ is generally multi-valued (when $V \gsim 0.6$). 
For example, two possible profiles with the same volume
$V=0.6$ (having $P_{\rm b}=3.86$ and 4.80,) are presented in 
the third column of Fig.~\ref{fig:shape_v05}.
Also, one can see from Fig.~\ref{fig:vu_pb05}(a) that there are many shapes
with the same volume, say $V=1.5$. 
Each column of Fig.~\ref{fig:shape_v05} presents several equilibrium 
profiles corresponding to a fixed volume.
The drop on the top of each column has the lowest bottom pressure
$P_{\rm b}$, and the second (third) one has the second (third) lowest 
$P_{\rm b}$. 
These profiles show that when the volume is fixed, the lower $P_{\rm b}$
corresponds to the shorter drop length (= the pressure difference 
between the top and the bottom of drop) as can easily be understood from
eq.~(\ref{eq:curvature}).

Observing Figs.~\ref{fig:vu_pb05}(a) and \ref{fig:vu_pb05}(b) together, 
one can see that
when the volume is fixed and if there are more than one equilibrium shapes
(which takes place for $V \gsim 0.6$), the shape with smaller 
$P_{\rm b}$, namely, a shorter shape, has a lower energy.
This means that a decrease of the surface energy due to lowering $P_{\rm b}$
overwhelms an increase of the gravitational energy induced by lifting up 
the center of mass.
The relation such that the lower $P_{\rm b}$ corresponds to
the lower $U$ generally holds in volume-radius limited systems but 
not for volume-angle limited ones like a drop
hanging from an infinite horizontal plane. 
(The volume-angle limited systems will be mentioned later on.)
Now one might expect that the stable shape has the smallest equilibrium 
energy.
But that is not self-evident.
In fact, it turns out that the statement is true for the volume-radius 
limited system, but not for the volume-angle limited one.
In order for any equilibrium shape to be realized, 
$U$ must increase 
under any infinitesimal deformation with the constraint such 
that both $V$ and $a$ are fixed.

In Appendix A, we derive a sufficient condition for stability.
Stability of shapes that do not satisfy this sufficient condition
are numerically examined and it was found 
that any equilibrium shape except the one 
having the lowest $P_{\rm b}$ (and hence the shortest one) is unstable 
under some axisymmetric deformation. 
That was examined for various values of the top radius $a$. 
On the other hand, it is difficult to determine whether or not the shortest shape, 
which has the lowest energy, is really stable 
under any infinitesimal deformation. 
We have studied several types of axisymmetric deformations
and conjecture that the shortest shape is stable at least under 
any axisymmetric deformation, so that the profile on the top of each column 
is realized when the volume is given as indicated. 
The drop on the right end in the first row, 
having a neck close to the tip of the faucet, is also
stable. Its volume $V=2.39$ is almost equal to the critical 
volume $V_{\rm c}$ indicated in Fig.~\ref{fig:vu_pb05}(a).
After all, as the volume increases, the profile of static drop changes 
from left to right in the first row of Fig.~\ref{fig:shape_v05}.
%%%%%%%%%%%%%%% fig.3 preprint
%\marginpar{\fbox{Fig.\ref{fig:shape_v05}}}
%%%%%%%%%%%%%%%

Interestingly, there are equilibrium shapes with infinite numbers of 
necks (for $V \approx 1.5$ when $a=0.5$), 
although such shapes are not realized because they are
unstable under infinitesimal deformations.

%%% 2.2 %%%%%%%%%%%%%%%%%%%%%%%%%%%%%%%%%%%%%%%%%%%%%%%%%
\subsection{Drop hanging from an infinite horizontal plane}

\begin{figure}[p]
%\figureheight{1cm}
	\begin{center}
	\includegraphics[width=1\linewidth]{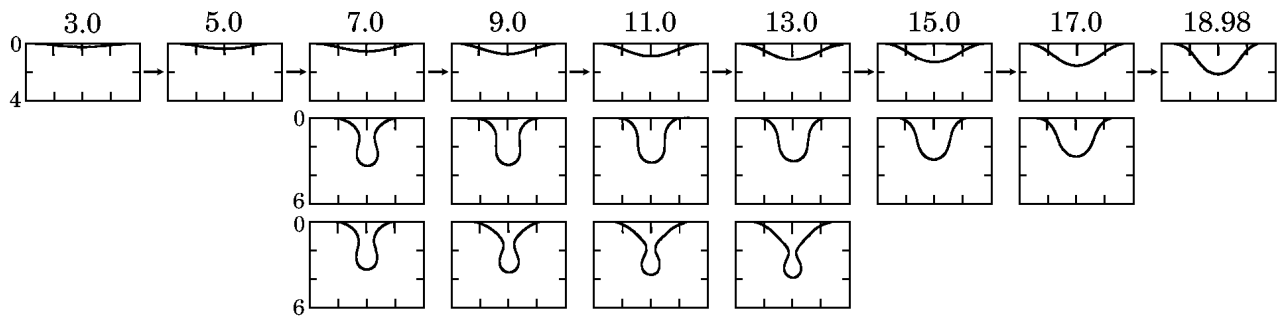}
	\end{center}
\caption{Same as in Fig. \ref{fig:shape_v05} \ but for a drop hanging 
from the ceiling.}
\label{fig:vu_pb}
\vspace{1cm}
\begin{center}
\end{center}
\end{figure}

\begin{figure}[p]
%\figureheight{1cm}
	\begin{center}
	\includegraphics[width=.5\linewidth]{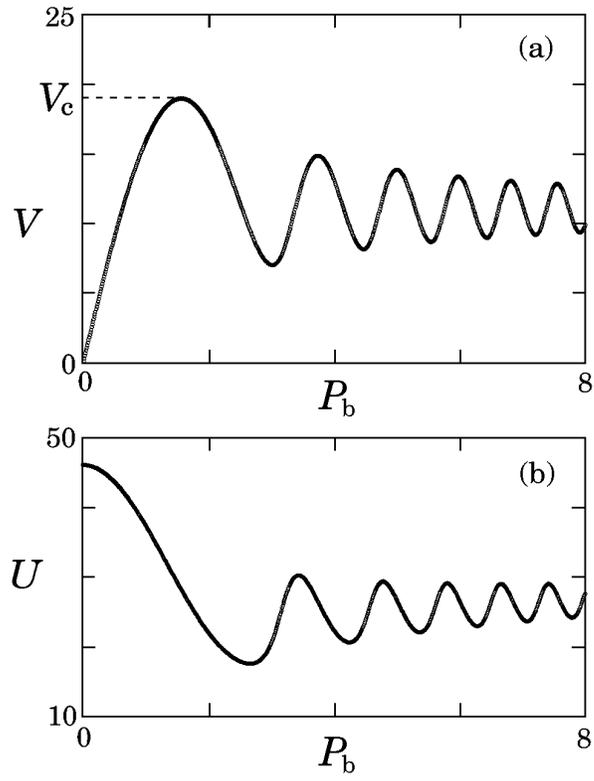}
	\end{center}
\caption{Same as in Fig. \ref{fig:vu_pb05} \ but for a drop 
hanging from the ceiling.
The maximum volume is $V=18.98$.}
\label{fig:shape_v}
\vspace{1cm}
\begin{center}
\end{center}
\end{figure}

Drops seen for example on the ceiling in the bath-room are volume-angle
limited systems. They can also be obtained
by integrating eq.~(\ref{eq:heikou1}) up to $\theta=\pi /2$ if the plane is
fully wetted.
Several equilibrium shapes with fixed volume
are presented in Fig.~\ref{fig:shape_v}. 
Plots of $V$ vs. $P_{\rm b}$ and $U$ vs $P_{\rm b}$ are given in Figs.~\ref{fig:vu_pb}(a) and \ref{fig:vu_pb}(b), respectively. 
In each column of Fig.~\ref{fig:shape_v}, profiles are 
arranged in increasing order of
$P_{\rm b}$ (and hence increasing order of drop length).
From a variational analysis, all profiles except the first row are found 
to be unstable under some axisymmetric deformation. As the volume increases,
the realized static shape of drop changes from left to right
in the top row. 

It should be noted that 
if there are more than one equilibrium shapes,
the stable one is the shortest 
but its energy is the second lowest as can be seen from
Figs.~\ref{fig:vu_pb}(a) and \ref{fig:vu_pb}(b).
%05Jan99 C
This is not surprising because the equilibrium state with the lowest energy 
is, as pointed out already, not necessarily stable for systems whose energy 
has not a lower bound. 
In the present case, each shape in the second top row of 
Fig.~\ref{fig:shape_v} has a lower energy than in the top row, 
which implies that a decrease of the gravitational energy overwhelms an increase of the surface energy. 
But the shape cannot be stable under some axisymmetric deformation.  

The critical volume is $V_{\rm c}\approx 18.98 (=0.37$cm$^3$ 
for water at 20$^{\circ}$ C).
A drop larger than this volume cannot be suspended from the ceiling.

%%%%%%%%%%%%%%%%%%%%%%%%%%%%%%%%%%%%%%%%%%%%%%%%%%%%%%%%%%%%%%%
%  3.Lagrangian description of fluid motion             %
%%%%%%%%%%%%%%%%%%%%%%%%%%%%%%%%%%%%%%%%%%%%%%%%%%%%%%%%%%%%%%%
\section{Lagrangian Description of Fluid Motion}
\setcounter{equation}{0}

We take the following assumptions:
\begin{enumerate}
\renewcommand{\labelenumi}{(\arabic{enumi})}
\item
Incompressible fluid.
\label{incomp}
\item
Axisymmetry.
\label{axisym}
\item
Horizontal component of the fluid velocity can be neglected 
in comparison with vertical one.
\label{vxy_neglect}
\item
Vertical component of the velocity depends only on the vertical 
coordinate.
\label{v_const}
\item
No exchange between upper and lower parts of fluid.
\label{no_exch}
\end{enumerate}
Assumption (\ref{no_exch}) is derived from assumptions (\ref{vxy_neglect}) 
and (\ref{v_const}).
These assumptions correspond to the shallow-water theory applied
to axisymmetric systems\cite{ll87} and have been used widely
\cite{lee74,schulkes94,eggers97}.

\begin{figure}[p]
%\figureheight{1cm}
	\begin{center}
	\includegraphics[width=.3\linewidth]{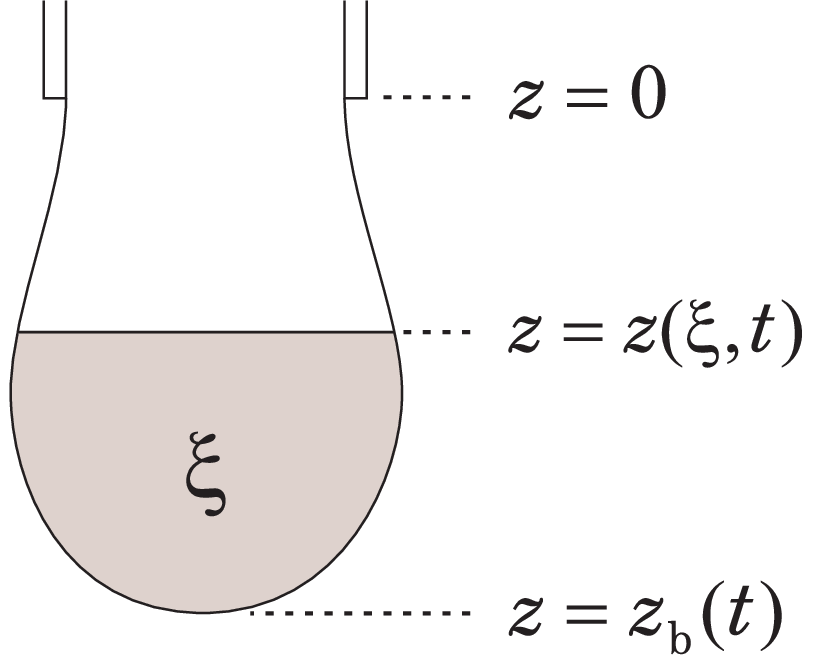}
	\end{center}
\caption{Definition of variables.}
\label{fig:xi_def}
\vspace{1cm}
\begin{center}
\end{center}
\end{figure}

Let us denote the volume of the fluid below the vertical 
coordinate $z$ 
(the positive direction is defined to be downward)
by
\begin{equation}
\xi(z,t) \equiv \int_z^{z_{\rm b}(t)} \pi r(\zeta,t)^2 {\rm d}\zeta \,,
\end{equation}
where $z_{\rm b}(t)$ is the vertical coordinate of the bottom at time $t$ and
$r(z,t)$ is the radius of the fluid at the coordinate $z$.
Then, from assumptions (\ref{no_exch}), 
$\xi$ can be used as a Lagrangian variable. 
(See Fig.~\ref{fig:xi_def}.) 

The kinetic energy is thus given as
\begin{equation}
E_{\rm kin} = \displaystyle\frac{\rho}{2}\int_0^{\xi_0(t)}
v^2 {\rm d}\xi\,,
\label{eq:ekin}
\end{equation}
\[
v \equiv v_z \equiv \frac{\partial z(\xi,t)}{\partial t}\,.
%\left\{ \frac{\partial z(\xi,t)}{\partial t} 
%\right\}^2 {\rm d}\xi\,.
\]
Here, 
the coordinate of the top of the drop ($=$ the end of the faucet)
is defined as $z=0$, and $\xi_0(t) \equiv \xi(0,t)$ is the total 
volume of the drop.
The gravitational energy is
\begin{equation}
U_g = -\rho g \int_0^{\xi_0(t)} z(\xi,t) {\rm d}\xi\,.
\label{eq:ug}
\end{equation}
The surface energy is expressed as
\begin{equation}
%\begin{array}{lll}
U_{\Gamma} = \Gamma 
\displaystyle\int_{0}^{z_{\rm b}(t)} 2 \pi r(z,t) 
\sqrt{1+ 
\{ \partial r (z,t)/ \partial z \}^2} {\rm d}z\,,
%\displaystyle\int_{z=0}^{z=z_{\rm b}(t)} 2 \pi r(z,t) {\rm d}s\,,
\label{eq:ugamma}
\end{equation}
%where
%\begin{equation}
%ds = \sqrt{1+ \{ \partial r (z,t)/ \partial z \}^2} dz\,.
%\end{equation}
which can be rewritten as
%$U_{\Gamma}$ can be rewritten as
\begin{equation}
U_{\Gamma} = \Gamma \displaystyle
\int_0^{\xi_0(t)}\sqrt{4\pi z' + 
\frac{(z'')^2}{(z')^4}} {\rm d}\xi\,,
\label{eq:ugamma1}
\end{equation}
where
\[
z' \equiv \displaystyle\frac{\partial z(\xi,t)}
{\partial \xi}\,, 
\hspace{1cm}
z'' \equiv \displaystyle\frac{\partial^2 z(\xi,t)}
{\partial \xi^2}\,. \]
Equations (\ref{eq:ekin}), (\ref{eq:ug}) and (\ref{eq:ugamma1}) yield 
Lagrangian of the system as
\begin{equation}
%\mbox{$\cal L$} = E_{\rm kin} - U_g - U_{\Gamma}
{\cal L} = E_{\rm kin} - U_g - U_{\Gamma}\,.
\label{eq:lagran}
\end{equation}
The effect of the viscosity is expressed by 
a dissipation function, 
%the following 
%energy dissipation,
$\dot{E}_{\rm kin}$ (namely, the time derivative of the Kinetic energy of fluid) as \cite{ll87}
\begin{equation}
\begin{array}{lll}
\dot{E}_{\rm kin} &=& -\displaystyle\frac{1}{2} 
\eta\int_0^{\xi_0(t)}\sum_{i,j} 
\left(\frac{\partial v_j}{\partial x_i} + 
\frac{\partial v_i}{\partial x_j} \right)^2 {\rm d}\xi \\
&=& -3\eta\displaystyle\int_0^{\xi_0(t)} 
\left\{\frac{\partial v(z,t)}{\partial z} \right\}
^2 {\rm d}\xi \,. 
\end{array}
\label{eq:edot}
\end{equation}
In the above, we have used the relation
\[
%{\rm div} \mbox{\boldmath $v$} = 
\displaystyle\frac{\partial v_x}{\partial x} 
+ \frac{\partial v_y}{\partial y} + 
\frac{\partial v_z}{\partial z} = 0\,,
\]
so that
\[
\displaystyle\frac{\partial v_x}{\partial x} = \frac{\partial v_y}{\partial y}
= -\frac{1}{2}\frac{\partial v_z}{\partial z}\,,
\]
because of assumptions (\ref{incomp}) and (\ref{axisym}).
%where $v(z,t)=v_z$ is the $z$-component of velocity at $z$.
Equation (\ref{eq:edot}) reads
\begin{equation}
\dot{E}_{\rm kin} = -3 \eta\displaystyle\int_0^{\xi_0(t)}
\frac{\{\partial v(\xi,t)/\partial\xi \}^2}
{\{\partial z(\xi,t)/\partial \xi \}^2} {\rm d}\xi\,.
\label{eq:dissipation}
\end{equation}

We discretize the integral forms 
(\ref{eq:ekin}),(\ref{eq:ug}),(\ref{eq:ugamma1}) and (\ref{eq:dissipation}).
Let the drop be sliced into $M$ disks by $(M-1)$ horizontal planes  
$z=z_1, z_2, \cdots z_{M-1}$ as Fig.~\ref{fig:dxi_def}.
The volumes of these disks, $\Delta\xi_j$, are expressed as
\begin{equation}
\Delta\xi_j = \int_{z_{j-1}}^{z_j} \pi r(\zeta, t)^2 {\rm d} \zeta; 
\quad \quad
j = 1, 2, \cdots M;
\end{equation}

\begin{equation}
z_M \equiv z_{\rm b}\,.
\end{equation} 
These volumes are conserved because of the incompressibility.
(Variables at the top and the bottom of the fluid are defined
suitably by taking account of the boundary condition. See below.)
The mass of the disks are
\[
\Delta m_j = \rho\Delta\xi_j\,,
\]
which reads
\begin{equation}
\Delta m_j = \Delta\xi_j
\end{equation}
by taking units such that $\rho=1$.
Let the coordinate of the center of mass of the $j$-th disk be
$\tilde{z}_j$. Then the kinetic and potential energies are given by
\begin{equation}
E_{\rm kin} \simeq \displaystyle\frac{1}{2}\sum_{j=1}^{M}\Delta m_j
(\dot{\tilde{z}}_j)^2, 
\label{eq:ekin2}
\end{equation}
\begin{equation}
U_g \simeq -g\sum_{j=1}^{M}\Delta m_j \tilde{z}_j.
\label{eq:ug1}
\end{equation}
The dissipation function is
\begin{equation}
\dot{E}_{\rm kin} \simeq -3\eta\sum_{j=1}^M\displaystyle
\frac{(v_j-v_{j-1})^2}
{(z_j-z_{j-1})^2}\Delta m_j,
\end{equation}
where
\[
v_j \equiv \dot{z_j} = \displaystyle\frac{\partial}{\partial t}z (\xi_j,t).
\]

The most important part of the algorithm is how to approximate the surface
energy because the rigorous expression of the surface tension force includes
higher-order derivatives in space coordinates.
We approximate the surface energy $U_{\Gamma}=\Gamma S$ ($S$ is the surface area of the fluid) by
patching many parts of conical surfaces: By  
defining the average radii of disks as
\[
r_j = \displaystyle\sqrt{\frac{\Delta m_j}
{\pi(z_j -z_{j-1})}}\,; 
\hspace{2cm}j=2, 3, \cdots M\,;
\]
the surface area in the interval
$[(z_{j-1} + z_j)/2, (z_j + z_{j+1})/2]$  is approximated as

%%%
%%%%%% 
%% preprint
%\begin{equation}
%S_j = \pi(r_j + r_{j+1})\sqrt{\displaystyle\frac{1}{4}
%(z_{j+1} + z_{j-1})^2 + (r_j - r_{j+1})^2}
%{\equiv} S_j (z_{j-1}, z_j, z_{j+1})\,. 
%%\hspace{2cm} j=1, 3, \cdots, M-1. \nonumber
%\end{equation}
\begin{eqnarray}
S_j &=& \pi(r_j + r_{j+1})\sqrt{\displaystyle\frac{1}{4}
(z_{j+1} + z_{j-1})^2 + (r_j - r_{j+1})^2} \nonumber \\
&{\equiv}& S_j (z_{j-1}, z_j, z_{j+1})\,. 
%\hspace{2cm} j=1, 3, \cdots, M-1. \nonumber
\end{eqnarray}
Then the surface energy becomes
\begin{equation}
U_{\Gamma} = \Gamma\sum_{j=1}^{M}S_j(z_{j-1}, z_j, z_{j+1})\,.
\end{equation}
Further, using an approximation  
\begin{equation}
\tilde{z}_j \simeq \displaystyle\frac{z_j + z_{j-1}}{2},
\hspace{0.5cm} 
\hspace{0.5cm} 
\dot{\tilde{z}}_j \simeq \displaystyle\frac{\dot{z}_j
+\dot{z}_{j-1}}{2}\,,
\label{eq:zj}
\end{equation}
and substituting eq.~(\ref{eq:zj}) into (\ref{eq:ekin2}) and (\ref{eq:ug1}),
we obtain a Lagrangian of the discretized system as
%%%
%%%%%%
%%%%%%%%%
%% preprint
%
%\begin{equation}
%%%\cal L \mit(z_1, \cdots z_N, \dot{z_1}, \cdots \dot{z_N}) =
%{\cal L}( z_1 , \cdots z_M, \dot{z}_1, \cdots \dot{z}_M) \nonumber \\
%={E_{\rm kin}}(\dot{z}_1, \cdots \dot{z}_M)
%- U_g(z_1, \cdots z_M)
%- U_{\Gamma}(z_1, \cdots z_M)\,. \nonumber \\
%\
%\end{equation}
\begin{eqnarray}
%\cal L \mit(z_1, \cdots z_N, \dot{z_1}, \cdots \dot{z_N}) =
{\cal L}(& z_1 &, \cdots z_M, \dot{z}_1, \cdots \dot{z}_M) \nonumber \\
&=&
{E_{\rm kin}}(\dot{z}_1, \cdots \dot{z}_M)
- U_g(z_1, \cdots z_M)
- U_{\Gamma}(z_1, \cdots z_M)\,. \nonumber \\
\
\end{eqnarray}

Equations of motion in the Lagrangian description are thus obtained from 
\begin{equation}
\displaystyle\frac{{\rm d}}{{\rm d}t}\frac{\partial {\cal L}}
{\partial \dot{z}_j}
= \frac{\partial{\cal L}}{\partial z_j} + \frac{1}{2}
\frac{\partial \dot{E}_{kin}}{\partial\dot{z}_j}\,; 
\hspace{0.5cm} j=1, 2, \cdots M.
\end{equation}
In the present simulation, we used a further approximation
\begin{equation}
\tilde{z}_j \simeq z_j, \hspace{1.0cm}
\dot{\tilde{z}}_j \simeq \dot{z}_j
\label{eq:approx_zj}
\end{equation}
in place of eq.~(\ref{eq:zj}).
%\vspace{0.5cm}

The top boundary is treated as follows:
We mark a part of the fluid just inside the faucet and let its 
coordinate be $z_0 ( <0$ ). The volume in 
the interval $[ z_0, z_1]$ is 
\begin{equation}
\Delta \xi_1 = \int_{z_0}^{z_1} \pi r(\zeta,t)^2 {\rm d}\zeta 
= \pi a^2|z_0| + \int_0 ^{z_1} \pi r(\zeta,t)^2 {\rm d}\zeta\,,
\end{equation}
which is constant as time passes. 
We fix the velocity of the fluid at the end of the faucet so that
\[
\dot{z}_0=v_0\,;
\]
namely, the volume of the fluid hanging from the faucet increases
steadily with the constant flow rare $Q= \pi a^2 v_0$.
The volume of the first disk i.e., the volume in the interval
[$0, z_1$], $\widetilde{\Delta \xi}_1$, increases as
\[
\widetilde{\Delta \xi}_1 =\int_0^{z_1}\pi r(\zeta,t)^2 {\rm d}\zeta
 =\Delta \xi_1 -\pi a^2|z_0|\,.
\]
The coordinate $z_1(\ge 0$) increases according to the above relation.
When 
%$z_1$ 
$\widetilde{\Delta \xi}_1$
reaches a certain constant value,
we redefine the numbers as
\begin{equation}
j+1 \leftarrow j\,;\hspace{0.5cm} 
j=1,2, \cdots M\,;
\end{equation}
and reset $z_0$ and $z_1$ to be 
\[
z_0 \leftarrow - \Delta \xi_1/\pi a^2, 
\quad \quad z_1 \leftarrow 0\,.
\]
In this way the number of disks $M$ increases by one.
$M$ also increases when the relative thickness of any disk exceeds a certain 
limit and is divided in two. 
A more detailed description of the algorithm will be given in Appendix B.

%%%%%%%%%%%%%%%%%%%%%%%%%%%%%%%%%%%%%%%%%%%%%%%%%%%%%%%%%%%%%%%
%  4.Results and discussions                                  %
%%%%%%%%%%%%%%%%%%%%%%%%%%%%%%%%%%%%%%%%%%%%%%%%%%%%%%%%%%%%%%%

\section{Results and Discussions}
\setcounter{equation}{0}

We choose $t_0 \equiv (\Gamma/ \rho g^3)^{1/4}$ as the time unit.  
For water at 20 $^{\circ}$C, $t_0=0.017$ s.  
The unit of viscosity was chosen as $\eta_0 \equiv 
(\rho \Gamma^3/g)^{1/4}$. Then  viscosity 
is $\eta=0.002$ for water at 20 $^{\circ}$C. 
(Other units have been defined as eq.~(\ref{eq:unit}).)
Setting values for the faucet radius $a$ and the 
velocity $v_0$, we 
simulate the time evolution of the dripping faucet.

%%% 4.1 %%%%%%%%%%%%%%%%%%%%%%%%%%%%%%%%%%%%%%%%%%%%%%%%%%
\subsection{Comparison of shapes with experiments}

Peregrine et al. observed details of the shapes of drops falling from
%the end of 
a capillary tube\cite{pss90}. We first see that the present algorithm can well 
reproduce their experimental data.
The faucet radius was chosen to be $a=0.952$ corresponding 5.2 mm in diameter 
in their experiment. 
The flow rate of the water is not mentioned explicitly except their
expression 'as slow as possible'. We chose as $v_0 = 
0.01(= 0.16$ cm/s).
This is the velocity at the top of the drop (i.e., the tip of the faucet)
employed as a boundary conditions. 
%so that the volume of the fluid
%increases linearly with the flow rate $Q=\pi a^2 v_0$ as time passes.
The breakup parameter is $\epsilon =10^{-4}$, which means
when the minimum cross section of the necking region divided by 
the cross section of the faucet, 
$S_{\rm min} \equiv (r_{\rm min}/a)^2$, 
reaches this value we separate the drop from
the remaining part and continue the simulation for the residue.

\begin{figure}[p]
%\figureheight{1cm}
	\begin{center}
	\includegraphics[width=1\linewidth]{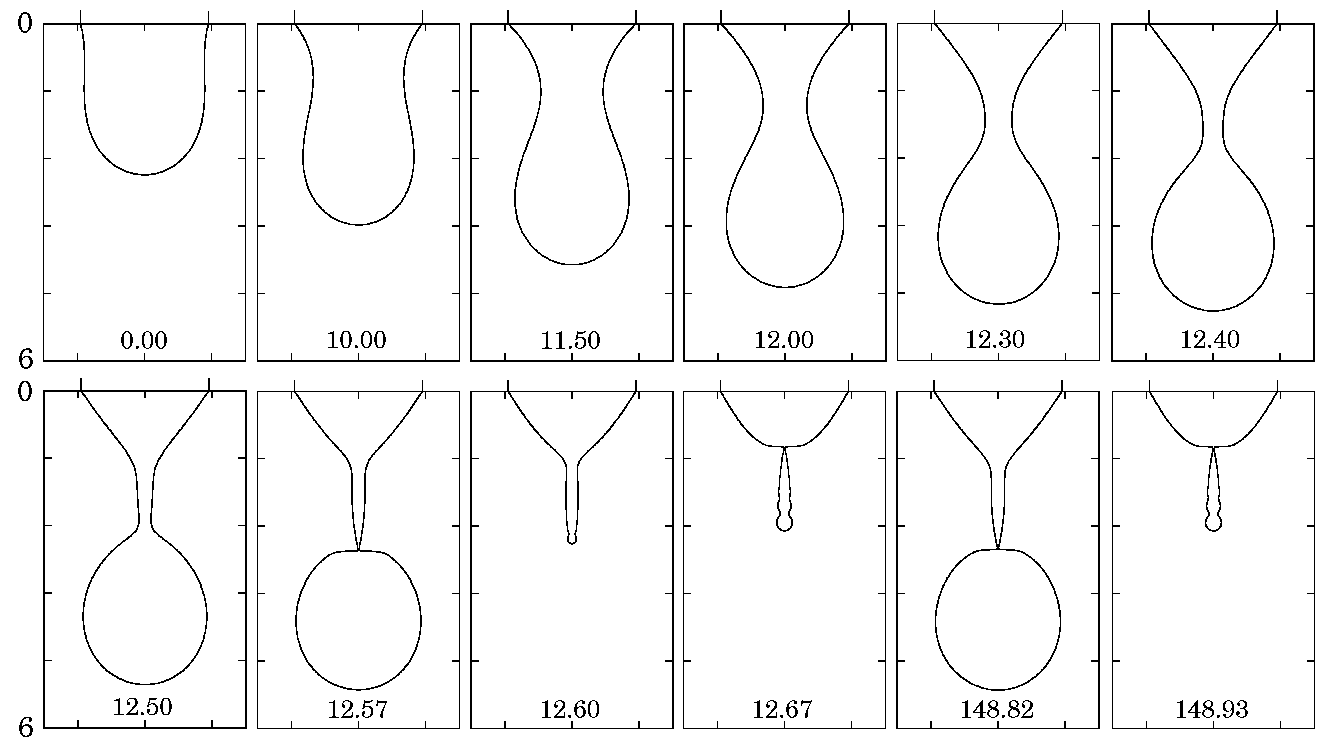}
	\end{center}
\caption{Temporal change of the shape.
$a=0.952$, $v_0=0.01$, $\eta=0.002$, $\epsilon=10^{-4}$.
Initial condition: $P_{\rm b}=2.6$ corresponding to
the initial volume $V_{\rm init}=4.77$.}
\label{fig:shape_t}
\vspace{1cm}
\begin{center}
\end{center}
\end{figure}

\begin{figure}[p]
%\figureheight{3cm}
	\begin{center}
	\includegraphics[width=1\linewidth]{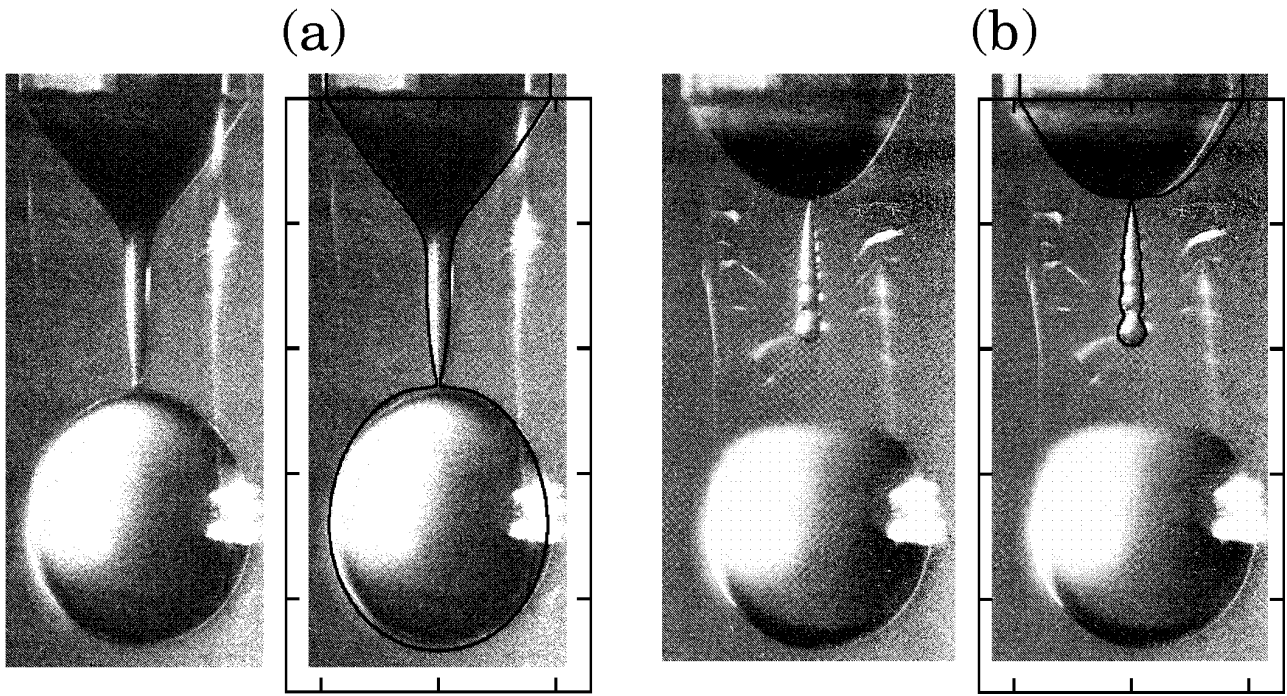}
	\end{center}
\caption{
A comparison of experimental and calculated drop shapes close to 
breakup moments. The photographs of the drops are from Peregrine et al.
(J. Fluid Mech.,{\bf 212} (1990) 25. Reprinted with permission
of Cambridge University Press.)
$a=0.952$, $v_0=0.01$, $\eta=0.002$, $\epsilon=10^{-4}$.  
(a) The moment at which $S_{\rm min}=5 \epsilon$.
(b) The moment at which $S_{\rm min}=\epsilon$.
}
\label{fig:compare}
\vspace{1cm}
\begin{center}
\end{center}
\end{figure}

In Fig.~\ref{fig:shape_t}, we present profiles of drop falling from the faucet at various stages of evolution.
The initial shape is taken from a static stable state of 
the volume $V_{\rm init}=4.77$, which
has been obtained by integrating eq.~(\ref{eq:heikou1}) with $P_{\rm b}= 2.6$.
The time $t=12.57$ is the breakup moment, namely the moment
at which $S_{\rm min}$ reaches $\epsilon$. The time $t=12.67$ is 
the second breakup moment.   
In Fig.~\ref{fig:compare}, the profiles near the breakup moment are
superimposed with the experimental photographs.

We see that the present simulation reproduces the observed shapes 
excellently.
Especially, our shape of the satellite (the secondary small droplet)
is closer to the observed one than that obtained from a different 
algorithm in which the viscosity is ignored.\cite{schulkes94} 
(The shape at the first breakup moment has been well reproduced by 
Eggers\cite{eggers97}.)
If we remember several approximations made in our algorithm
(particularly, eq.~(\ref{eq:approx_zj}) and the procedure (\ref{division}) 
in Appendix B), 
%it is rather amazing that 
the agreement between simulation and experiment is amazingly good.

We can estimate an error of the breakup moment based on the following
scaling relations\cite{eggers97,eggers97a}:
\begin{eqnarray}
r(t) & \propto & (t_{\rm c}-t)^{2/3} \quad {\rm for 
\quad inviscid \quad fluid}\,,
\label{eq:inviscid} \\
r(t) & \propto & (t_{\rm c}-t)^1  \qquad {\rm for \quad viscid \quad fluid}\,,
\label{eq:viscid}
\end{eqnarray} 
where $t_{\rm c}$ is the critical point.
The scaling region where eq.~(\ref{eq:viscid}) holds is 
extremely small for water due to small viscosity
(typical scales are
$l_{\nu} \equiv \eta^2/\rho \Gamma \sim 1.4 \times 10^{-6}$cm, 
$t_{\nu} \equiv \eta^3/\rho \Gamma^2 \sim 1.9 \times 10^{-10}$s).  
On the other hand, eq.(\ref{eq:inviscid}) holds only approximately
because of a crossover effect\cite{eggers97a}.
Nevertheless, these relations are useful enough to estimate 
the upper and lower limits of $t_{\rm c}$.
Using eq.~(\ref{eq:inviscid}) and eq.~(\ref{eq:viscid}) to 
extrapolate the plot of $r(t)$, we obtained 
, for example, $12.58 < t_{\rm c} < 12.66$ corresponding to 
the first breakup moment $t=12.57$.

%%% 4.2 %%%%%%%%%%%%%%%%%%%%%%%%%%%%%%%%%%%%%%%%%%%%%%%%%%%
\subsection{General feature of drops falling from a faucet}

\begin{figure}[p]
%\figureheight{1cm}
	\begin{center}
	\includegraphics[width=.4\linewidth]{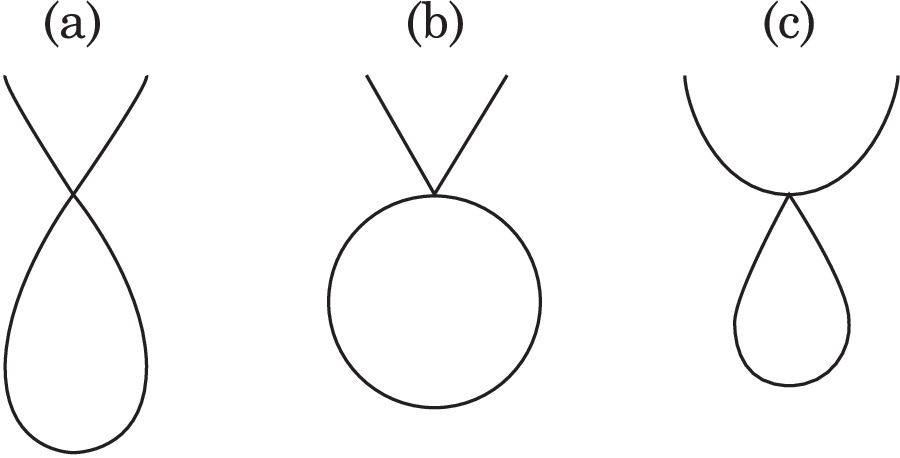}
	\end{center}
\caption{The shape at the critical point is like (b) or (c) instead of (a).
(c) is for pinching of a satellite.}
\label{fig:symmetry}
\vspace{1cm}
\begin{center}
\end{center}
\end{figure}

As has already been recognized experimentally \cite{pss90} and 
%confirmed
theoretically 
\cite{schulkes94,eggers97}, 
%\cite{schulkes94,ed94}, 
liquid drops do not have 
up-down symmetry like Fig.~\ref{fig:symmetry}(a) near 
the critical point at which a drop separates. 
That is not due to the gravity. Instead, the symmetry spontaneously 
breaks down as sketched in Fig.~\ref{fig:symmetry}(b) or (c) 
even without the gravitational effect. 
(The shape (c) corresponds to breakup of a satellite.)  
As we have seen, the marked asymmetry
is observed in the present simulation as well. 
It should be noted that
the asymmetry is a dynamical property of the fluid with free surface,
therefore it cannot be predictable from the equilibrium shapes as in 
Fig.~\ref{fig:shape_v05} and also not reproduced in the
simulation neglecting kinetic terms\cite{op93}.  

\begin{figure}[p]
%\figureheight{1cm}
	\begin{center}
	\includegraphics[width=.15\linewidth]{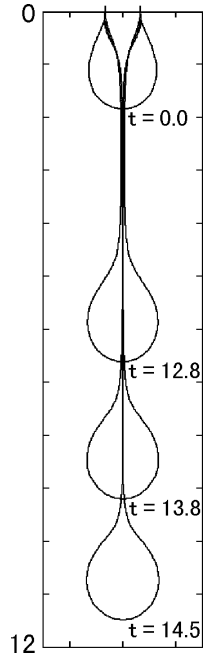}
	\end{center}
\caption{Time evolution of a drop of glycerol. 
The length, time and viscosity units are $l_0 = 0.266$ cm, 
$t_0 = 0.0125$ s and $\eta _0 = $4.26 g/cm s, respectively, 
corresponding to glycerol at 20 $^{\circ}$C.~ $a = 0.332$ (1.5 mm diameter), 
$\eta=3.50$, $v_0 = 0.05$. }
\label{fig:hi_visco}
\vspace{1cm}
\begin{center}
\end{center}
\end{figure}

Theoretically, from scaling properties derived by Eggers
\cite{eggers93,eggers97},
the asymmetric shapes near the critical point are expected to be
universal irrespective of the viscosity, the faucet radius and the flow rate
except the inviscid case.
Practically however, it is also well known 
%both experimentally and theoretically 
that the global shapes are strongly dependent on these conditions because
the scaling law, even if it is exact, works only in the small region 
of time and space
close to the critical point. 
In fact, the liquid bridge that connects the conical part (just below the
faucet) and the spherical part shows a rough tendency to become 
long and thin for (1) large $\eta$, (2) large $a$ and (3) small
$v_0$. 
These features are confirmed also in our simulation.
%05Jan99 D
For example, we present a simulation for large $\eta$. 
Figure \ref{fig:hi_visco} shows a temporal change of a drop of glycerol 
corresponding to an experiment by Shi et al \cite{sbn94}. 
Similar profiles have been obtained by solving the Navier-Stokes 
equations \cite{ed94}. 
Both results well reproduce a long bridge observed in the experiment.

\begin{figure}[p]
%\figureheight{1cm}
	\begin{center}
	\includegraphics[width=.4\linewidth]{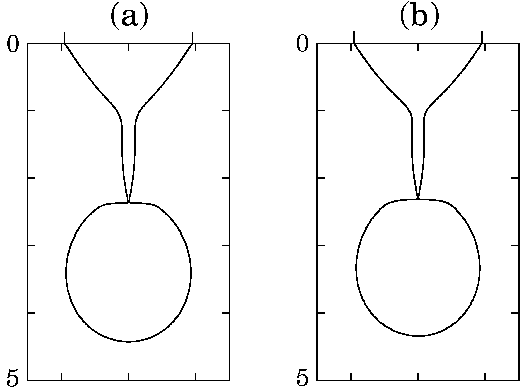}
	\end{center}
\caption{Profile at the critical point in the limit of $v_0 \rightarrow 0$.
$a=0.952$, $\eta=0.002$, $\epsilon=10^{-4}$.
(a) $v_0=0.01$. (b) $v_0=0$.}
\label{fig:v0_dep}
\vspace{1cm}
\begin{center}
%Nobuko {\sc Fuchikami} , Shunya {\sc Ishioka} and Ken {\sc Kiyono}
\end{center}
\end{figure}

Strictly speaking, 
the length of the liquid bridge does not change monotonically
with these parameters 
because of spatial and temporal oscillations of
the liquid surface.  
The necking region is delicately affected by the interplay 
of these oscillations and liquid influx.
Concerning the third condition (i.e., $v_0$-dependence), 
the shape tends to a final
one in the limit of $v_0 \rightarrow 0$ which should coincides with that
obtained by setting $v_0 =0$ and starting from an initial state with
volume just above the critical value $V_{\rm c}$.
For example, in case of parameter values ($\eta=0.002$, $a=0.952$, 
$v_0=0.01$, 
$\epsilon=10^{-4}$), with the initial volume $V_{\rm init} =4.77$,
the first breakup occurred at the drop size
$V_{\rm d}=3.85$ and the residue $V_{\rm r}=1.28$ 
(Fig.~\ref{fig:v0_dep}(a), which is the same profile 
at $t=12.57$ in Fig.~\ref{fig:shape_t}).
These results are to be compared with $V_{\rm d}=3.72$ and $V_{\rm r}=1.23$,
which were obtained by a simulation with no flow: $v_0=0$, the same
values for other parameters and the initial volume $V_{\rm init}=4.95$ 
slightly 
larger than the critical value
%$V_{\rm c}$
(Fig.~\ref{fig:v0_dep}(b)).
Two profiles look almost the same.

%%% 4.3 %%%%%%%%%%%%%%%%%%%%%%%%%%%%%%%%%%%%%%%%%%%%%%%%%%%%%%%
\subsection{Formation of secondary drop}

\begin{figure}[p]
%\figureheight{1cm}
	\begin{center}
	\includegraphics[width=.5\linewidth]{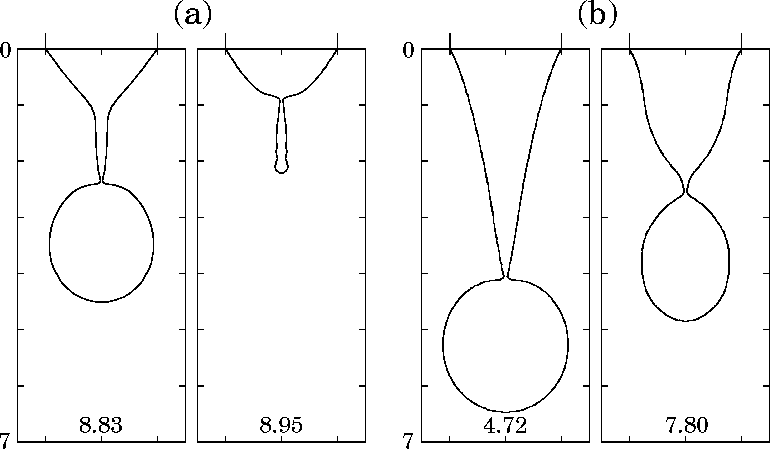}
	\end{center}
\caption{Formation of the secondary drop. $a=1.0$, $\eta=0.002$, 
$\epsilon = 10^{-3}$. 
Initial condition $P_{\rm b}=2.6$ corresponding to $V_{\rm init}=5.21$.
Profiles are at the breakup moment, at which 
$S_{\rm min}= \epsilon$. The time of the breakup moment is written in each
frame.
(a) $v_0=0.003$. (b) $v_0=0.3$.}
\label{fig:subdrop}
\vspace{1cm}
\begin{center}
\end{center}
\end{figure}

In the experiments focusing on long-term behavior of dripping faucets,
the flow rate is usually chosen as a control parameter.
As mentioned above, the shape of drop is strongly affected by the velocity $v_0$ or, equivalently, the flow rate.
In Figs.~\ref{fig:subdrop}(a) and \ref{fig:subdrop}(b), 
we present how the shape of drop at the second breakup moment
depends on $v_0$.
Parameters are chosen as 
($a=1.0$, $\eta=0.002$, $\epsilon = 10^{-3}$) commonly in (a) and (b); 
$v_0=0.003$ in (a) and $v_0=0.3$ in (b).
The initial shape was commonly taken to be a static stable one.
In both (a) and (b), two profiles describe the first and the second 
breakup moments (i.e., $S_{\rm min}= \epsilon$).

For smaller $v_0$, after the spherical drop detached from the
bottom of the neck, a satellite droplet forms from the slender neck itself
and separates from the conical part just below the faucet.
The satellite is thus much smaller than the main drop. 
The volume ratio is 
%less than 1 \% 
0.7 \% in the present example. 
For larger $v_0$, the volume increases so rapidly that 
the slender liquid bridge disappears, the spherical part and the 
cone jointing directly. 
In the latter case, the secondary drop is again spherical and relatively
large, 
its size being 44 \% of the main drop.
Results for the secondary-drop formation similar to Fig.~\ref{fig:subdrop}
have been obtained from the simulation of inviscid fluid\cite{schulkes94}.

\begin{figure}[p]
%\figureheight{3cm}
	\begin{center}
	\includegraphics[width=1\linewidth]{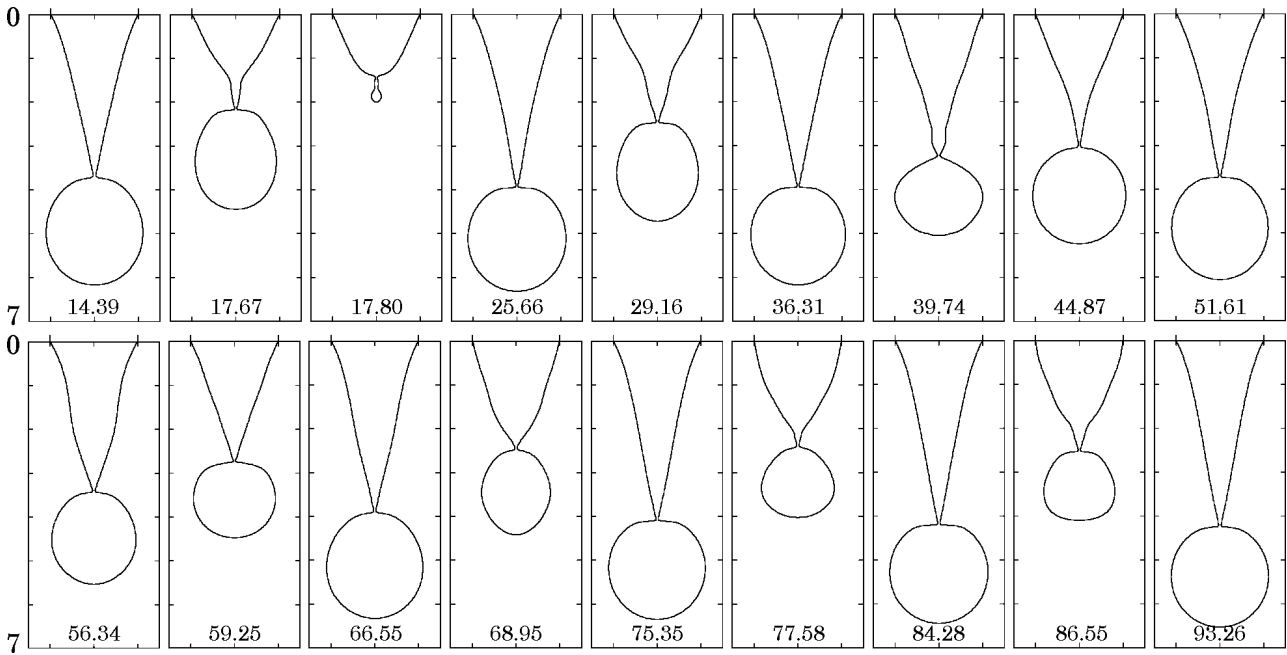}
	\end{center}
\caption{Various shapes at each moment of breakup, at which 
$S_{\rm min} =\epsilon$. Continuation of Fig.\ref{fig:subdrop}(b). 
$a=1.0$, $v_0=0.3$, $\epsilon =10^{-3}$.}
\label{fig:series_chig}
\vspace{1cm}
\begin{center}
\end{center}
\end{figure}

Now a question is whether such successive processes, breakup of a main
large drop followed by a smaller secondary drop can 
take place regularly or not.
Details of shapes after the second breakup moment has not been reported 
so far. 
Figure \ref{fig:series_chig} 
represents a continuation of Fig.~\ref{fig:subdrop}(b), namely, 
a series of breakup moments 
for the same parameter values: 
$a=1.0, \; v_0=0.3$, where the first profile represents
the third breakup moment.  
%($S_{\rm min}=\epsilon$) 
%for the same parameter values as Fig.~\ref{fig:subdrop}(b): 
%$a=1.0, \; v_0=0.3$, while 
%$\epsilon=10^{-3}$ was used to decrease computational time. 
There appears sometimes a tiny droplet (smaller than $\sim 0.01$ \% of 
the largest drops)
splitting from the tip of the liquid cone just after the main drip,  
that is not presented here. 
%For smaller value of $\epsilon$, the size of 
%these droplets becomes even smaller.
Computational error may have led to such small
droplets but the possibility that extremely small droplets really
appear cannot be excluded. 
A larger drop and a smaller drop appear alternatively in 
Fig.~\ref{fig:series_chig}, which looks almost periodic.
Note, however, that large (small) drops are slightly different
from each other in size and shape. These fluctuations are surely
intrinsic, not due to computational error.
%Obviously, 
%Many variables describing the system cannot be 
%exactly the same at each breakup moment. 
%at which only $S_{\rm min}$ has the same value $\epsilon$. 
A minute difference at each breakup moment can be an origin of 
not only fluctuations but also an irregular motion under
certain conditions.
Here we see that the dripping faucet is really a complex system.
%%%%%%%%%%%%%%%% fig.13 preprint
%\marginpar{\fbox{Fig.\ref{fig:series_chig}}}
%%%%%%%%%%%%%%
%%% 4.4 %%%%%%%%%%%%%%%%%%%%%%%%%%%%%%%%%%%%%%%%%%%%%%%%%%%%%%
\subsection{Long-term behavior}

We present here results for $a=0.916$ corresponding to the faucet of 5mm
in diameter used in a recent
experiment by Katsuyama and Nagata\cite{kn98}.
For the long-term simulation, a larger value for the breakup parameter 
$\epsilon =4 \times 10^{-3}$ was used to save the computational time.
The velocity $v_0$ is chosen to be relatively small ($<0.16$), so that
the secondary drop formation is like Fig.~\ref{fig:subdrop}(a)
instead of Fig.~\ref{fig:subdrop}(b). 
In other words, pinching off of a satellite 
(or occasionally two successive satellites) 
%the size of which is 
smaller than 1 \% of the largest drops always occurs just after 
breakup of a main drop.
%some short time less than $\sim 0.2$ 
%($\lsim 0.2$) after each breakup of the main drops. 
In the following analysis, we ignore
these satellites. 
%together with extremely tiny droplets mentioned already
Then, the drop sizes
%the sizes of the main drops 
distribute mostly in the range larger than 10 \% of the largest drops.  

\begin{figure}[p]
%\figureheight{3cm}
	\begin{center}
	\includegraphics[width=1\linewidth]{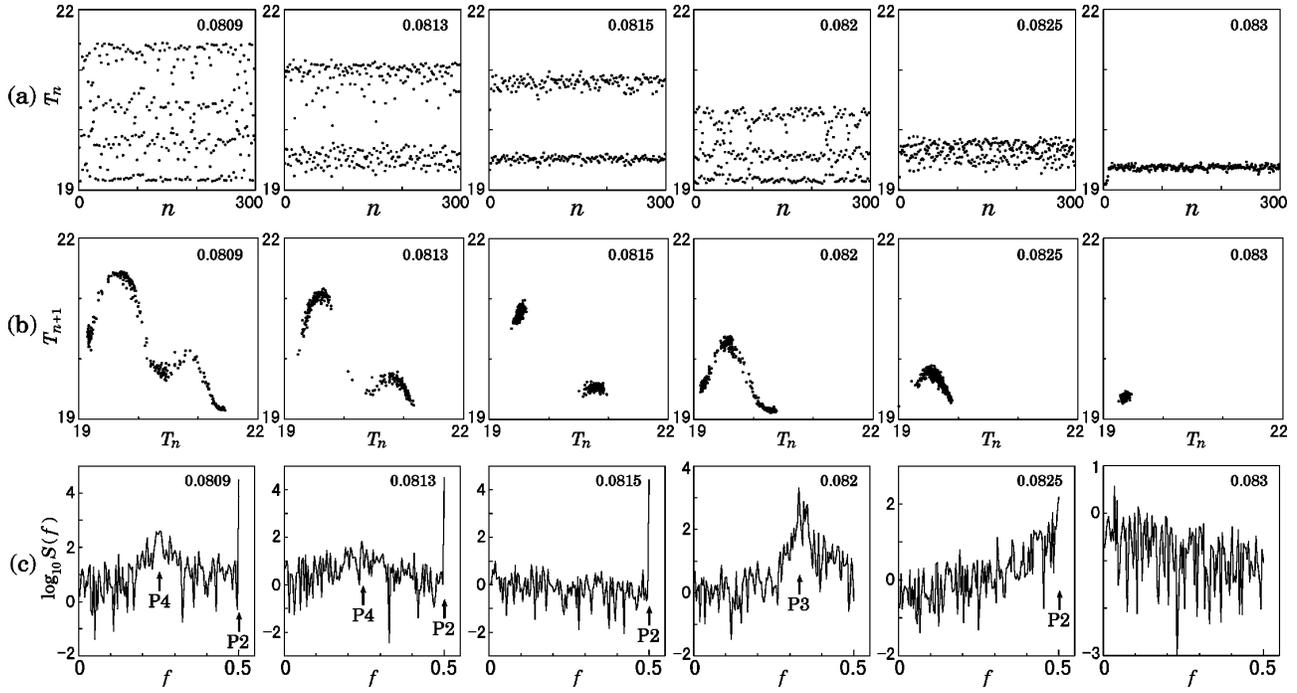}
	\end{center}
\caption{
(a) Plot of dripping time interval $T_n$ vs $n$ for various
values of $v_0$. Value of $v_0$ is written in each frame.
(b) Return map: plot of $T_{n+1}$ vs $T_n$.
(c) Semi-log plot of power spectrum of (a) calculated from
$2^{8}$ data points.
}
\label{fig:series916_v0}
\vspace{1cm}
\begin{center}
\end{center}
\end{figure}

We let $T_n$ denote the $n$-th dripping interval, i.e., the time difference
between the $(n+1)$-th and $n$-th drips.
Neglect of satellites above mentioned implies that intervals $T_n$ 
which are smaller than $\sim 0.2$ are omitted, while 
$T_n$ for main drops are larger than $\sim 10$. 
Figure \ref{fig:series916_v0}(a) represents plotting of $T_n$ vs $n$ for
various values of $v_0$. Corresponding return maps, i.e., plots of
$T_{n+1}$ vs $T_n$ and power spectra are given in Figs.~\ref{fig:series916_v0}
(b), and (c), respectively.

The time series data $\{T_n\}$ fluctuate considerably, but the power spectra
can suggest types of bifurcation.
As the control parameter 
$v_0$ is varied, 
%decreases, 
period one (P1) motion at $v_0=0.083$ period-doubles $backward$ to 
period two (P2) motion. That can be confirmed from 
% ($v_0$=0.0825). 
the power spectrum for $v_0$=0.0825 which exhibits 
a peak at frequency $f=1/2$, indicating P2 oscillation. 
When $v_0=0.082$, intermittent period three (P3) motion is observed, 
where the spectrum has a sharp peak at $f=1/3$. 
Another type of P2 motion is observed at $v_0=0.0815$. We found that
%this type of P2 motion is not the one 
the bifurcation to this type of P2 motion is $not$ period doubling 
from any P1 motion. 
Rather, forward and backward period doubling cascades to chaos, 
starting from this type of P2 motion,
are expected. In fact, the spectra for $v_0=0.0813$ and 0.0809
exhibit a peak at $f=1/4$ besides a sharp peak at $f=1/2$,
which indicates that the P2 motion at $v_0=0.0815$ period-doubles backward.
The backward period doubling is widely observed also experimentally.
Attractors in Fig.~\ref{fig:series916_v0}(b) closely resemble 
experimental ones.

\begin{figure}[p]
%\figureheight{3cm}
	\begin{center}
	\includegraphics[width=.5\linewidth]{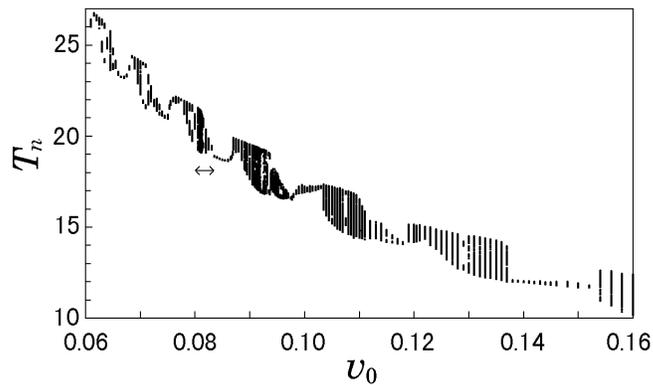}
	\end{center}
\caption{
Bifurcation diagram of the dripping interval $T_n$ vs $v_0$.
}
\label{fig:bif916_v0}
\vspace{1cm}
\begin{center}
\end{center}
\end{figure}

Figure \ref{fig:bif916_v0} is a bifurcation diagram in which the parameter 
range of Fig.~\ref{fig:series916_v0} is indicated.   
As suggested by Fig.~\ref{fig:bif916_v0}, oscillations like those shown in 
Fig.~\ref{fig:series916_v0} are quite typical in the present
system, and repeatedly appear at different values of $v_0$.  
Similar types of oscillations are observed not only in the experiment by Katsuyama and Nagata for the same faucet size
\cite{kn98}, but also in other experiments with 
smaller faucets\cite{dh91,sgp94}.    

In the present simulation, at $v_0=0.074$ and $v_0 =0.083$ for instance, 
the $T_n$ value is uniqe. 
In contrast, between these $v_0$ values (for $0.074<v_0<0.083$ for instance), 
the $T_n$ value is distributed over a finit range. 
This is seen on Fig.~\ref{fig:bif916_v0} in the form of a bloc, 
which repeats as $v_0$ increases. 
We may call this pattern a unit structure. 
Looking at Fig.~\ref{fig:bif916_v0} with reference to $T_n$, 
one sees each unit structure occurs periodically, namely, 
at the same interval in $T_n$.  
Qualitative agreement of our bifurcation diagram 
with the experiment by Katsuyama and Nagata is satisfactory in a wide range of
the control parameter $v_0$, or equivalently the flow rate.
In their bifurcation diagram, a unit structure similar to ours also appears repeatedly as the flow rate is varied.

\subsection{Further study}

We have seen that the present simulation reproduces various aspects of the 
dripping faucet system which are observed experimentally.
%, i.e., return maps,
%bachward bifurcation, intermittency, unit structure of the bifurcation diagram
%etc.
The next question is why the dripping faucet behaves like this.
We will answer to this question in the forthcoming paper.\cite{kf99}
Here we give only an outline of it:

We construct an improved mass-spring model based on a detailed analysis of our
numerical simulation. The model reveals the basic mechanism of the complex
behavior of the dripping faucet system.
A key of the new model is that the mass dependence of the spring-constant
is taken into account. 
A similar global feature observed
in the experiment and the simulation, i.e., the repeating unit structure
in the bifurcation diagram, is reproduced also in the improved mass-spring
model.
This simplified model shows that each unit in the bifurcation diagram 
includes ordinary period doubling cascade to chaos, 
forward and backward period doubling cascade starting from P2 motion, 
intermittency and hysteresis.
It is also clarified that the unit structure 
can be explained in terms of
oscillations of the center of mass of the fluid, in other words, 
each unit is characterized by an integer which 
relates to the frequency of the oscillations during
the time between two successive breakup moments.
%These results will be reported in the next paper.

%\section*{Acknowledgement}
\section*{Acknowledgment}

We wish to acknowledge valuable discussions with Dr T. Katsuyama.  
We thank him and Professor K. Nagata for showing us their
data prior to publication. 
We are grateful to Professor M. Inokuti for critical reading of 
the manuscript and many helpful comments.
%Professor M. Inokuti  
%read a part of our manuscript carefully and made many corrections
%in it, which we really appreciate.
Thanks are also due to Professor D. H. Peregrine and
Cambridge University Press for permission  
to copy photographs appearing in the article 
``The bifurcation of liquid bridges''
by D. H. Peregrine, G. Shoker and A. Symon in
J. Fluid Mech. {\bf 212} (1990) 25.

%We are grateful to Dr. Katsuyama for valuable discussions. 
%We thank him and Professor Nagata for showing us their
%data prior to publication. 
%We would like to thank Professor Inoguchi for critical reading of the manuscript and helpful comments.  
%We also appreciate Professor Peregrine and
%Cambridge University Press for permission  
%to copy photographs appearing in the article 
%``The bifurcation of liquid bridges''
%by D. H. Peregrine, G. Shoker and A. Symon in
%J. Fluid Mech. {\bf 212} (1990) 25.

%%%%%%%%%%%%%%%%%%%%%%%%%%%%%%%%%%%%%%%%%%%%%%%%%%%%%%%%%%%%%%%
%  Appendix A                                                 %
%%%%%%%%%%%%%%%%%%%%%%%%%%%%%%%%%%%%%%%%%%%%%%%%%%%%%%%%%%%%%%%
\appendix
\section{Stability Analysis}

We consider an axisymmetric deformation such that the part between the
interval $[z, z+{\rm d}z]$ is mapped onto the interval $[Z, Z+{\rm d}Z]$, where
$$
{\rm d}Z=\left[1+\epsilon (z)\right]{\rm d}z
$$
with the constraint
\begin{equation}
\epsilon(0)= \epsilon(z_{\rm b})=0\,.
\label{eq:constraint}
\end{equation}
The coordinates at the top and the bottom of the drop are defined as 
$z=0$ and $z_{\rm b}$, respectively, which are mapped onto
$Z=0$ and $Z_{\rm b}$ by the deformation.

Because the volume of each part is conserved before and after the deformation,
the radius $r(z)$ is transformed as 
$$
R(Z)=\frac{r(z)}{\sqrt{1+\epsilon (z)}}\,.
$$ 
The increment of the surface energy $U_{\Gamma}$ caused by 
the deformation is expressed as
\begin{eqnarray}
\delta U_{\Gamma} & \equiv & 2 \pi \Gamma \int_0^{Z_{\rm b}}
R \sqrt{1+(R')^2} {\rm d}Z \nonumber \\
& & {} - 2 \pi \Gamma \int_0^{z_{\rm b}}
r \sqrt{1+(r')^2} {\rm d}z \nonumber \\
&=& \delta U_{\Gamma,1} + \delta U_{\Gamma,2}\,, \nonumber
\end{eqnarray}
where $\delta U_{\Gamma,1}= O(\epsilon)$ and
$\delta U_{\Gamma,2} =O(\epsilon^2)$ are the first and the second order
quantities of the small deformation $\epsilon$.
The increment of the gravitational energy $U_{g}$ is
$$
\delta U_{g} \equiv - \rho g \int_0^{Z_{\rm b}}
\pi R^2 Z {\rm d}Z + \rho g  \int_0^{z_{\rm b}}
\pi r^2 z {\rm d}z = \delta U_{g,1};
$$
$\delta U_g$ includes only linear terms of $\epsilon$,
namely the second order term vanishes: $\delta U_{g,2} =0$.
It is a little tedious but not difficult to derive the force-balance 
equation (\ref{eq:heikou}) and (\ref{eq:curvature}) from
the equilibrium condition
$$
\delta U_{\Gamma,1}+\delta U_{g,1}=0.
$$

The stability condition is
$$
\delta U_{\Gamma, 2}>0,
$$
where
\begin{eqnarray}
\delta U_{\Gamma,2} & = &
\frac{\pi \Gamma}{4} \int_0^{z_{\rm b}}
\left[ \phi(z)\epsilon^2 + \psi(z) (\epsilon')^2 \right]{\rm d}z\,, \\
\phi(z) & \equiv & \frac{r}{\left(1+(r')^2\right)^{5/2}} \nonumber \\
& & {} \times \left[ -1 + (r')^2 + (r')^4 - r r'' \left(7-2 (r')^2
\right)\right]\,, 
\label{eq:phi_z}
\nonumber \\
\ \\
\psi(z) & \equiv & \frac{r^3}{\left(1+(r')^2\right)^{3/2}}\,.
\end{eqnarray}
The second term in the integrand is positive definite. 
Therefore the condition
\begin{equation}
\phi(z)>0 \quad {\rm for} \quad 0<z<z_{\rm b}
\label{eq:sufficient}
\end{equation}
is sufficient for the stability of 
equilibrium shape under any axisymmetric deformation $\epsilon(z)$. 

%However, the above condition is too strong.
When the drop has a neck, the condition (\ref{eq:sufficient}) 
is not satisfied because $\phi(z)$ is negative at the neck 
($r'=0$, $r''>0$). We have found that 
all shapes with volume $V\gsim 1.3$ in 
Fig.~\ref{fig:shape_v05} and $V \gsim 17.0$ in Fig.~\ref{fig:shape_v}
violate this sufficient condition. 
%even when 
(Some of them have no neck but do not satisfy (\ref{eq:sufficient}).)
However, such shapes are not necessarily unstable because the
the homogeneous deformation 
$\epsilon ' =0$ is impossible under the constraint
of eq.~(\ref{eq:constraint}), and hence the second term of the integrant in
eq.~(\ref{eq:phi_z}) always   
works in the direction of stabilization.

%However, the above condition is too strong 
%because the homogeneous deformation: $\epsilon=$constant so that 
%$\epsilon ' =0$ is impossible under the constraint
%of eq.(\ref{eq:constraint}) and then the second term always 
%works in the direction on stabilization.
%For example, all shapes with the volume $V\gsim 1.3$ in 
%Fig.~\ref{fig:shape_v05} and $V \gsim 17.0$ in Fig.\ref{fig:shape_v} cannot 
%satisfy the above condition. Among them, the case when the shape has a neck
%(as the largest volume in Fig.\ref{fig:shape_v05}) we easily see the
%violation from this condition because 
%$\phi(z)$ becomes negative around the necking point at which
%$r'=0$.
Therefore we first calculate $\phi(z)$ and if
$\phi(z)<0$ in an interval $0 \le z_1 < z < z_1 + d$, 
the following infinitesimal deformation is considered:
\begin{eqnarray}
\epsilon(z) & = & 2 \Delta \frac{(z-z_2)}{(\alpha d)^2 \beta}
\quad \nonumber \\ 
& & \hspace{1cm} {\rm for} \quad z_2 \le z \le \beta \alpha d + z_2,\\ 
\epsilon(z) & = & 2 \Delta \frac{(\alpha d +z_2 -z)}{(\alpha d)^2 (1-\beta)}
\quad \nonumber \\ 
& & \hspace{1cm} {\rm for} \quad \beta \alpha d + z_2 \le z \le \alpha d +z_2,\\
\epsilon(z) & = & 0 \quad{\rm otherwise},
\end{eqnarray}
which leads to a stretch of the original shape by the length 
$$
\int_{z_2}^{z_2+\alpha d} \epsilon {\rm d}z = \Delta\,.
$$
The parameters $z_2$ (satisfying $0\le z_2<z_2+ \alpha d<z_{\rm b}$),
$\alpha$ ($>0$), and $\beta$ ($>0$) characterizing the deformation
are varied and we examined if $\delta U_{\Gamma,2}/\Delta^2$ becomes negative.
In this way, for fixed values of $V$ and $a$, 
all the equilibrium shapes except the shortest one, 
which has the lowest energy, are found to be
unstable for the volume-radius limited systems.
Also for the volume-angle limited systems, 
all the equilibrium shapes except the shortest one, 
which has the second lowest energy,
are found to be unstable.  
All the shapes (including the largest one) on the top of  
each column in Figs.~\ref{fig:shape_v05} and \ref{fig:shape_v} are found 
to be stable at least under axisymmetric deformations.

%%%%%%%%%%%%%%%%%%%%%%%%%%%%%%%%%%%%%%%%%%%%%%%%%%%%%%%%%%%%%%%
%  Appendix B                                                 %
%%%%%%%%%%%%%%%%%%%%%%%%%%%%%%%%%%%%%%%%%%%%%%%%%%%%%%%%%%%%%%%
\section{Algorithm}

We employed a fourth-order Runge-Kutta method
with adaptive stepsize control\cite{press97}.
When the radius $a$ of the faucet and the velocity $v_0$ of liquid at 
the tip of the faucet are given, the simulation is done based on 
the following algorithm.   

\begin{figure}[p]
%\figureheight{1cm}
	\begin{center}
	\includegraphics[width=.4\linewidth]{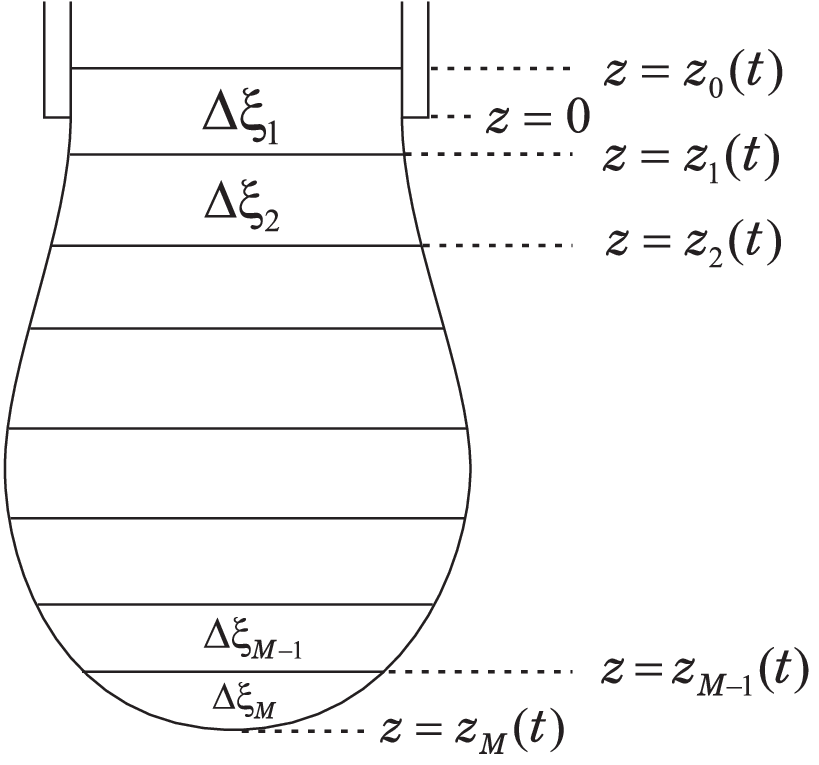}
	\end{center}
\caption{Definition of variables.}
\label{fig:dxi_def}
\vspace{1cm}
\begin{center}
\end{center}
\end{figure}

\begin{enumerate}
\renewcommand{\labelenumi}{(\arabic{enumi})}
\item
Initial shape\\
\label{initial}
We choose one of the stable equilibrium shapes obtained from the method
described in \ref{sec:heikou}.

\item
Decomposition of a drop\\
The drop of (\ref{initial}) is sliced with horizontal planes so that the drop is 
decomposed into many disks. The thickness of disks are such that the length
$s$ of each disk measured along the edge of the section shown in 
Fig.~\ref{fig:dxi_def} is equal. Therefore 
the disks near the bottom are 
relatively thin(see Fig.~\ref{fig:dxi_def}). 
The total number of the disks is denoted as $M$.

\item
Coordinates and volumes of disks\\
Vertical coordinates of the slice plane are denoted as
$z_1,\; z_2,\; \cdots z_{M-1}$. They satisfy the relation
\begin{equation}
0 \le z_1<z_2< \cdots <z_M \equiv z_{\rm b}\,,
\label{eq:order}
\end{equation}
where
\[
z_1(t=0)=0\,.
\]
Further, we set $z_0$ to be a negative constant value:
$z_0=z_0(t=0)<0$.
The volume of the disk in the interval $[z_{j-1}, z_j]$ are denoted as 
$\Delta \xi_j (j=1,2,\cdots M)$  and the volume in the 
interval $[0, z_1]$ as $\widetilde{\Delta \xi}_1$.
These volumes are calculated precisely from the static solution employed as 
an initial condition.

\item
Average radius of disks\\
We define the average radii of disks as follows:
\begin{eqnarray}
\left.
\begin{array}{llll}
r_j &=& \sqrt{\displaystyle\frac{\Delta\xi_j}
{\pi(z_j -z_{j-1})}} & \equiv \quad r_j(z_j, z_{j-1})\,; 
\\
%& & & \hspace{0.2cm}; \hspace{0.2cm}  j=2,3,\cdots M\,; \\ 
& & & j=2,3,\cdots M\,; \\
r_1 &=& \sqrt{\displaystyle\frac{\widetilde{\Delta\xi}_1}
{\pi z_1}}\,.  & \hspace{0.2cm} (z_1 >0)
\end{array}
\right\} \nonumber \\
\
\label{eq:rj}
\end{eqnarray}

\item
Initial velocities of disks\\
We set as
$v_j(t=0) = v_0$ ($j=1,2, \cdots v_M$).

\item
Tentative stepsize\\
At $t=0$, we tentatively set as 
$\Delta t = \widetilde{\Delta t}$.

\item
Adaptive stepsize\\
\label{adaptive_step}
From the time-evolution equations
\begin{eqnarray}
\left.
\begin{array}{lll}
\dot{z}_j &=& v_j\,,\\
\dot{v}_j &=& f_j (\{z_i\},\{r_i(z_i,z_{i-1})\},\{v_i\}) \,;\\
& &  \hspace{0.5cm} j=1,2,\cdots M\,;
\end{array}
\right\} \nonumber \\
\
\label{eq:jikan}
\end{eqnarray}
we estimate the error $\widetilde{\Delta}$ for one step of fourth-order 
Runge-Kutta algorithm. Then $\widetilde{\Delta}$ is compared with
our desired accuracy $\Delta_0$ and an adjusted stepsize $\Delta t$ is decided.
This procedure is applied for all errors of variables
$\{z_i\},\{v_i\}$ and the smallest value of $\Delta t$ is taken
\cite{press97}. 

\item
Time evolution\\
\label{time_evolution}
Equation (\ref{eq:jikan}) is integrated by one time step i.e., from $t$ to
$t+\Delta t$ using the fourth-order Runge-Kutta method.
\label{}
\item
Renewal of the stepsize\\
From (\ref{time_evolution}), 
we estimate the error $\widetilde \Delta$ in a similar way
to (\ref{adaptive_step}) and renew the adapted
stepsize $\Delta t$ for the next step.

\item
Check of overtaking\\
Because of assumption (\ref{no_exch}) in \S 4, any disk must not overtake the
neighboring one, in other words, the relation (\ref{eq:order}) should be maintained all times. So if the condition of eq.~(\ref{eq:order})
is violated, we choose a smaller value for $\Delta t$ and redo the
same procedure (\ref{time_evolution}).
This does not occur if the accuracy $\Delta_0$ has been chosen small enough.

\item
Increment of the number of disks\\
When the volume of the first disk, $\widetilde{\Delta \xi}_1$ 
becomes larger than a certain constant value,
we redefine variables as
\[
\begin{array}{lll}
z_{j+1} &\leftarrow& z_j\,; \quad j=1,2,\cdots M\,;\\
v_{j+1} &\leftarrow& v_j\,; \quad j=0,1,\cdots M\,;\\
z_1 &\leftarrow& 0\,, 
\end{array}
\]
and reset $z_0$ to be the initial value $z_0(t=0)$.
In this way, $M$ increases by one.

\item
Division of a disk\\
\label{division}
If the ratio of the width to the radius,
$(z_j - z_{j-1})/r_j$ is over a certain limit 
for any disk, the disk is divided into two parts so as to 
conserve both the total volume and the momentum. In this way, 
$M$ is increased by one. It should be noted that the way of division 
is not unique. On the other hand, it turns out that there is no solution 
of division under additional constraint such that the energy
is rigorously conserved. We divided disks in such a way that the spatial 
derivatives of velocity is conserved besides the volume and the momentum. 

\item
Coalescence of disks\\
When the radii of some neighboring disks become larger than a certain value,
the two disks are combined into one. Then $M$ decreases by one.

\item
Renewal of disk radii\\
From the renewed values of $\{z_i\}$ and $\{\Delta\xi_i\}$, the values of
$\{r_i\}$ are replaced using eq.~(\ref{eq:rj}).

\item
Breakup of the drop\\
When the shape of the drop has a neck (or necks), 
the minimum radius $r_{\rm min}$ in the necking region is compared with 
the faucet radius.
If its relative cross section $S_{\rm min} \equiv (r_m/a)^2$ is less 
than a critical value,
then the part below the neck is separated. After that, the 
number of disks becomes $M \leftarrow (m-1)$ where $m$ is the disk number
corresponding to $r_m = r_{\rm min}$.

\item
Return to (\ref{time_evolution})\\
We go back to the procedure (\ref{time_evolution}) and continue 
the iterations.

\end{enumerate}

%%%% preprint
\newpage

%%%%%%%%%%%%%%%%%%%%%% References %%%%%%%%%%%%%%%

%%%15
%%\begin{figure}
%%\figureheight{3cm}
%%\caption{Plot of the volume vs dripping time.}
%%\label{fig:internittency}
%%\end{figure}

\renewcommand\figurename{Fig.}

%%%%%%%%%% fig.1

%\clearpage

%%%%%%%%%%%%%%% fig.2  exchangerd  08/jan/99

%\clearpage

%%%%%%%%%% fig.3  exchanged  08/jan/99 

%\clearpage
%%%%%%%%%%%%%%% fig.4

%\clearpage
%%%%%%%%%%%%%%% fig.5

%\clearpage
%%%%%%%%%%%%%%% fig.6

%\clearpage
%%%%%%%%%%%%%%%% fig.7

%\clearpage

%%%%%%%%%%%%%%% fig.8

%\clearpage

%%%%%%%%%%%%%%% fig.9

%\clearpage
%%%%%%%%%%%%%%%% fig.10

%\clearpage
%%%%%%%%%%%%%%%% fig.11 05Jan99

%\clearpage
%%%%%%%%%%%%%%%% fig.11

%\clearpage
%%%%%%%%%%%%%%%%% fig.12

%\clearpage
%%%%%%%%%%%%%%%% fig.13

%\clearpage
%%%%%%%%%%%%%%%%%%%% fig.14

%\clearpage
%%%%%%%%%%%%%%%%%%%% fig.15

\end{document}